\documentclass[conference]{IEEEtran}
\IEEEoverridecommandlockouts

\usepackage[utf8]{inputenc}
\usepackage[T1]{fontenc}
\usepackage{xurl} 
\usepackage{cite}
\usepackage{amsmath,amssymb,amsfonts}
\usepackage{algorithmic}
\usepackage{graphicx}
\usepackage{textcomp}
\usepackage{xcolor}
\usepackage{booktabs}

\def\BibTeX{{\rm B\kern-.05em{\sc i\kern-.025em b}\kern-.08em
    T\kern-.1667em\lower.7ex\hbox{E}\kern-.125emX}}
\begin{document}
\raggedbottom

\title{Snippet-Driven Supply Chain Discovery with LLMs: Scaling Visibility in China}

\author{
\IEEEauthorblockN{Hiroto Fukada}
\IEEEauthorblockA{
\textit{Graduate Institute for Advanced Studies (SOKENDAI)} \\
\textit{National Institute of Informatics} \\
Tokyo, Japan \\
ORCID: 0009-0009-2126-4842}
\and
\IEEEauthorblockN{Takayuki Mizuno}
\IEEEauthorblockA{
\textit{National Institute of Informatics} \\
\textit{Graduate Institute for Advanced Studies (SOKENDAI)} \\
Tokyo, Japan\\
ORCID: 0000-0003-1332-076X}
}

\maketitle

\begin{abstract}
Financial and economic research often relies on structured supply-chain disclosures and commercial databases. In China, supplier--customer disclosure is typically limited to major partners of listed firms, leaving unlisted firms and long-tail inter-firm links poorly captured in structured data. Public web evidence can partly complement this gap through corporate, government, and trade-media disclosures; however, full-text web mining at scale is costly because pages are often inaccessible or expensive to process with large language models (LLMs).
We propose a snippet-driven method for constructing a supply chain knowledge graph (SCKG), with firms as nodes and inter-firm relationships as edges. Web search snippets are query-biased summaries returned with search results. We use them as a scalable first-pass evidence layer for LLM-based relationship extraction.
We evaluate the pipeline in terms of extraction efficiency and coverage. For extraction efficiency, exhaustive full-text chunking discovers 19.8$\times$ more unique relationships than snippets, but requires 251.2$\times$ more input tokens and yields higher redundancy. For coverage, we use 130,685 Chinese firms as search seeds, covering Shanghai/Shenzhen-listed firms and large unlisted firms as of 2024. In the listed-firm subset, the resulting SCKG covers 7.2$\times$ more firms and 9.3$\times$ more relationships than the CSMAR disclosure-based benchmark, while revealing heavy-tailed degree patterns. Retained provenance metadata make the SCKG an auditable complement to disclosure-based databases.
\end{abstract}


\section{Introduction}
Supply chain (SC) networks are fundamental conduits through which economic shocks propagate, inducing fragility and abrupt fluctuations in broader economic systems~\cite{elliott2022networks,Carvalho2020SupplyChainDisruptions}. 
For investors, these linkages are not merely operational details but financially material information. Economic links between customers and suppliers have been shown to predict stock returns when investors underreact to news about connected firms~\cite{Cohen2008EconomicLinks}. More generally, information asymmetry can distort investment and financing decisions, while higher-quality disclosure is associated with more efficient investment~\cite{Biddle2009FinancialReportingQuality}. Recent evidence from China further suggests that supply-chain information disclosure is associated with corporate investment efficiency and that forward-looking SC risk disclosure can reduce information asymmetry in equity issuance~\cite{Gao2023SCDisclosureInvestmentEfficiency,Li2023SCDisclosureSEO}.
 
Despite this financial importance, global SC structures remain fragmented. Constructing comprehensive data infrastructures to capture SC topology has been recognized as an urgent international priority, motivating calls for interdisciplinary alliances~\cite{Pichler2023Alliance}. China occupies a central position in this landscape as one of the world's leading manufacturing powers; the high dependency of G7 nations on Chinese production renders Chinese SCs a critical hub within global networks~\cite{baldwin2023hidden}. Nevertheless, the visibility of Chinese SCs remains coarse~\cite{Shi_Yang_Li_2019,su17072828}. Improving the resolution of Chinese SC data is therefore important not only for policy analysis, but also for financial risk assessment and investment decision-making under incomplete information.
 
A central obstacle is not the absence of any data, but the institutional boundary of observable structured data. Existing structured sources include listed-firm annual reports, stock-exchange filings, government disclosure systems, and commercial financial databases that compile these records. For Chinese supplier--customer research, China Stock Market \& Accounting Research Database (CSMAR) is a useful and widely used disclosure-based benchmark because it compiles annual-report information for listed firms~\cite{CSMAROfficial}. However, the underlying disclosure regime typically reports only major partners, such as the top-five suppliers and customers, and therefore leaves unlisted firms and long-tail trading relationships outside the structured observation window~\cite{Huan2017TopFiveCustomers,Lin2021SupplyChainDiversificationCOVID,Cheng2022SupplierConcentrationInvestors}. We therefore treat CSMAR as a high-quality but disclosure-constrained benchmark, not as a complete ground truth for the underlying economy.

Public web evidence can partly complement this boundary because supply-chain links often become visible through decentralized disclosure incentives. Suppliers may advertise major customers to signal credibility, local governments may announce industrial partnerships or procurement projects, firms may publish implementation cases and partner lists, and trade media may report production, investment, or partnership events. In other words, even when a firm does not disclose a relationship directly, the relationship may appear from the supplier side, customer side, government side, or media side. This makes the web fragmented and noisy, but potentially valuable as an evidence layer for relationships outside formal disclosure databases.
 
Recent studies have explored the use of LLMs for constructing supply chain knowledge graphs (SCKGs) because LLMs can transform unstructured web text into structured relationship records without requiring domain-specific training data~\cite{AlMahri2026,WangTsung2025CASE_AKG_SC,Wadhwa2023RevisitingRE}. A knowledge graph (KG) represents entities as nodes and semantic relationships as edges~\cite{Ji2022KGSurvey}. In this paper, an SCKG is a firm-level KG in which nodes represent companies, directed edges represent supplier--customer or related trading relationships, and edge attributes store relation type, exchanged product or service, supporting evidence, source information, retrieval metadata, and credibility labels. Existing LLM-based SCKG approaches, however, rely primarily on full-text web documents or pre-existing structured datasets as information sources. Full-text retrieval is subject to paywalls, \texttt{robots.txt} restrictions, login requirements, and HTTP access failures, while document-scale LLM processing is expensive when targeting tens of thousands of firms.
 
We propose leveraging \emph{web search snippets}---short query-dependent descriptions displayed with search results---as the primary information source for broad first-pass SC relationship extraction. Search engines generate snippets primarily from page content and sometimes from metadata, and site owners can control whether and how snippets appear~\cite{GoogleSearchCentralSnippets}. Thus, snippets should not be interpreted as unrestricted access to paywalled or blocked documents. Their practical advantage is operational: one search API call returns multiple ranked results, each with a title, URL, and query-biased summary, providing compact evidence from many candidate pages without requiring our crawler to download, parse, and process each full document. Query-biased summaries have been shown to efficiently convey document content relevant to information needs~\cite{tombros1998advantages}. Our method combines this search-result evidence with LLM-based relation extraction, source-domain credibility labeling, and entity resolution to construct an auditable SCKG.
 
We validate our approach through experiments on Chinese SCs, addressing three research questions:
\begin{enumerate}
  \item[\textbf{RQ1.}] \textbf{Snippet vs.\ Full-Text:} Are snippets a practical and cost-efficient first-pass data source for SC relationship extraction compared with chunked full-text processing?
  \item[\textbf{RQ2.}] \textbf{SC Coverage:} How much additional firm and relationship coverage does the snippet-derived SCKG provide relative to the CSMAR disclosure-based benchmark, and does its degree distribution exhibit empirical regularities observed across inter-firm transaction networks?
  \item[\textbf{RQ3.}] \textbf{Credibility Assessment:} What fraction of the extracted network remains under progressively stricter source-credibility filters, and how can retained source metadata provide an auditable basis for post-hoc review?
\end{enumerate}

The contributions of this paper are threefold. First, we propose and validate a scalable data-collection framework that extracts SC relationships from search-result snippets without requiring full-text access to every source page.
Second, we apply the pipeline to 130,685 Chinese target firms, covering both Shanghai/Shenzhen-listed firms and large unlisted firms as of 2024, and show that the listed-firm subset substantially expands visibility beyond CSMAR's major-partner disclosure boundary. 
Third, we retain source URLs, snippet identifiers, and domain credibility labels as edge-level provenance, providing an auditable basis for post-hoc review.

\section{Related Work}
\subsection{Web-Based Information Extraction}
 
Web text has long been recognized as a domain-free and effective source for open information extraction~\cite{etzioni2008open}. 
More recently, its utility has been extended to SC relationship extraction specifically. 
Ristoski et al.~\cite{ristoski2020large} demonstrated large-scale relation extraction from web documents combined with knowledge graphs, incorporating human-in-the-loop validation. 
AlMahri et al.~\cite{AlMahri2026} showed that web-sourced information can enhance SC visibility when processed through LLMs. However, practical access to full-text web content is severely constrained by paywalls, \texttt{robots.txt} policies, and HTTP errors; 
in AlMahri et al.'s own experiments, the data sources were limited to Wikipedia, highlighting the difficulty of scaling full-text collection to tens of thousands of firms.
 
\subsection{LLM-Based SCKG Construction}
 
AlMahri et al.~\cite{AlMahri2026} proposed using LLMs' relation extraction capabilities~\cite{Wadhwa2023RevisitingRE} on full-text news articles and similar sources to improve SC visibility. 
Wang and Tsung~\cite{WangTsung2025CASE_AKG_SC} further explored automated KG construction for SC datasets assisted by LLMs. These studies are important because they show that LLMs can convert unstructured textual evidence into structured supply-chain relationships and thereby reduce the dependence on manually curated databases. However, their reliance on full-text access limits practical applicability at scale.

\begin{figure*}[t]
  \centering
  \includegraphics[width=\textwidth]{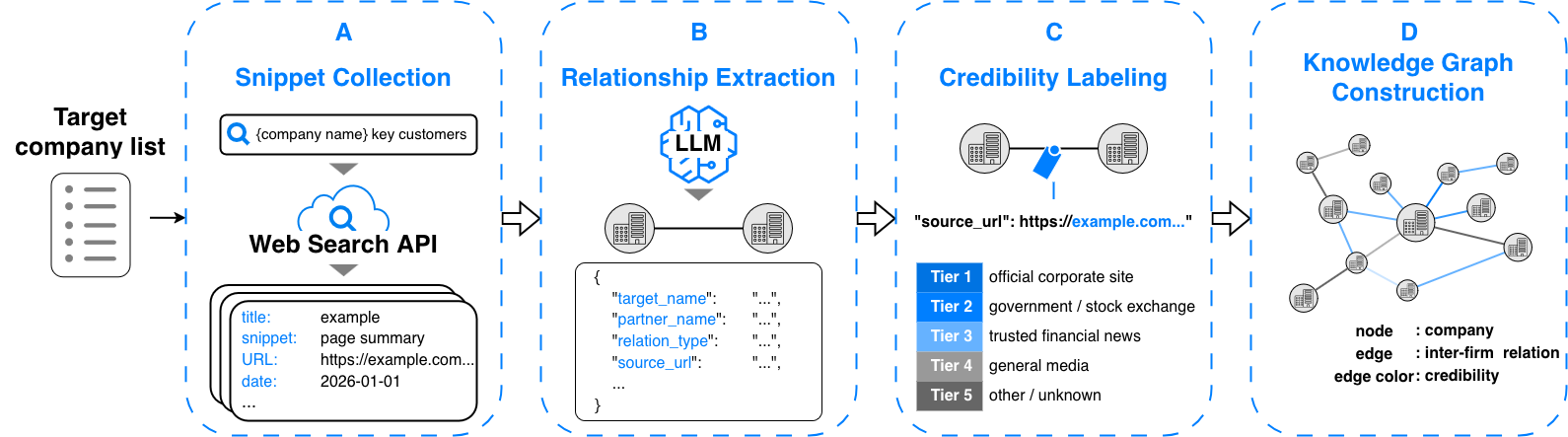}
  \caption{Overview of the proposed snippet-driven SCKG construction pipeline.}
  \label{fig:architecture}
\end{figure*}

\subsection{Our Positioning}

Our work differs from prior web-based SC extraction studies along three dimensions: information source, scalability, and auditability. 
First, we use web search snippets as compact evidence units rather than assuming full-text access to every source page. Snippets are query-biased summaries that efficiently convey relevant document content~\cite{tombros1998advantages}; using the title, URL, and summary text returned with ranked search results allows the pipeline to collect evidence from multiple candidate pages per query. 
Second, snippet-based extraction reduces dependence on full-text crawling and document-scale LLM processing, making it suitable for broad first-pass screening across a large target universe. 
Finally, our design retains per-snippet source URLs, snippet identifiers, and source credibility labels as edge-level provenance, enabling post-hoc review and filtering of extracted relationships.

\section{Method}

Figure~\ref{fig:architecture} summarizes the proposed pipeline. Starting from a target company list, the pipeline consists of four modules: (A) snippet collection through a web search API, (B) LLM-based relationship extraction into a structured JSON schema, (C) domain-based credibility labeling, and (D) entity-resolved SCKG construction. The design objective is to transform compact search-result evidence into an auditable first-pass supply-chain graph.

The target company list defines the set of focal firms for
which search queries are issued, as shown in Fig.~\ref{fig:architecture}. It is designed to cover both firms within the listed-firm disclosure boundary and firms outside that boundary, which are absent from disclosure-based datasets. 

\subsection{Snippet Collection}
 
A search snippet is the short, query-dependent text shown with a search result to help users judge whether a page is relevant. Search engines generate snippets primarily from page content and metadata, and site owners can restrict snippet display through mechanisms such as \texttt{nosnippet} or \texttt{max-snippet}~\cite{GoogleSearchCentralSnippets}. We therefore treat snippets as compact search-result summaries, not as unrestricted access to the underlying pages. Their benefit for this study is operational: a single search call returns multiple ranked pages with titles, URLs, and query-biased summaries, providing localized evidence for candidate relationship extraction while reducing dependence on full-text fetching and parsing.

Snippet collection provides standardized, traceable input records for each target firm. For each company, we issue predefined queries to a web search API and store the target identifier, query, retrieval timestamp, title, snippet text, source URL, publication date if available, and result-page/rank information. These fields correspond to the search-result records shown in the first module of Fig.~\ref{fig:architecture}.
 
\subsection{Relationship Extraction}

Relationship extraction converts snippet-level textual evidence into structured firm-to-firm edges. The model receives the target company name, page titles, and numbered snippet texts, and returns JSON fields for partner name, relation type, exchanged product or service, and evidence references. Source URLs and snippet identifiers are retained from the input records and attached to each extracted edge during post-processing. This step corresponds to the second module of Fig.~\ref{fig:architecture}, where unstructured snippet sets are transformed into structured output records.
 
\subsection{Credibility Labeling}

Credibility labeling makes large-scale discovery reviewable rather than treating every web source as equally reliable. We assign a credibility tier to each snippet's source domain through a three-phase decision process:

\begin{itemize}
  \item \textbf{Phase~1 --- Whitelist matching:} 
  The domain is checked against curated whitelists. 
  Government or stock exchange domains (e.g., \texttt{sse.com.cn}, \texttt{gov.cn}) yield \textbf{Tier~2} (official disclosure). 
  Trusted financial news domains (e.g., \texttt{eastmoney.com}, \texttt{bloomberg.com}) yield \textbf{Tier~3}.
  \item \textbf{Phase~2 --- Fuzzy matching:} 
  The domain is compared against the target and partner company names (augmented with an alias table) using fuzzy string matching. 
  A match assigns \textbf{Tier~1} (official corporate site).
  \item \textbf{Phase~3 --- Keyword matching:} 
  Domains containing news-related keywords (e.g., ``news,'' ``media,'' ``press'') are classified as \textbf{Tier~4} (general media). 
  All remaining domains default to \textbf{Tier~5} (aggregators, forums, or unknown).
\end{itemize}

The tier is not interpreted as a precision estimate. Instead, it records source-type information as edge-level metadata, enabling downstream review to be prioritized by source type, as represented in the third module of Fig.~\ref{fig:architecture}.
 
\subsection{Knowledge Graph Construction}
 
We construct a firm-level knowledge graph in which nodes represent firms and edges represent trading relationships enriched with metadata such as relation type, traded goods, textual evidence, source URLs, snippet identifiers, retrieval timestamps, and credibility tiers. This structure makes the extracted relationships analyzable as a network while preserving the provenance needed for downstream audit.

Entity resolution is necessary because the same firm may appear under multiple surface forms in web evidence. We first remove relationships whose relation type is \textsc{Unknown}, since they cannot be interpreted as supply-chain edges. Entity names are then normalized by removing legal suffixes (e.g., ``Inc.,'' ``Chinese limited-company suffixes'') and lowercasing English names.

We match entities in three stages: exact matching on normalized names, exact matching through a multilingual alias knowledge graph constructed from Wikidata and Orbis, and Jaro--Winkler similarity matching for remaining unmatched names. Wikidata helps link Chinese legal names, English names, abbreviations, former names, and alternative spellings of the same firm~\cite{Vrandecic2014Wikidata}. Orbis contains company records for the target firms used in the large-scale experiment, including Shanghai/Shenzhen-listed firms and large unlisted firms\cite{MoodysOrbis}. Unmatched entities are added as new nodes, yielding the SCKG shown in the fourth module of Fig.~\ref{fig:architecture}.

\section{Experimental Setup}

\subsection{Datasets}

Our experiments combine one disclosure-based benchmark network with two auxiliary resources. CSMAR provides the benchmark supplier--customer network for Chinese listed firms, while Orbis and Wikidata are used only for target-firm selection and entity resolution, not as sources of extracted supply-chain edges.

CSMAR covers China's stock markets and listed-company financial statements~\cite{CSMAROfficial}. We use its supplier--customer records for mainland-listed firms as the benchmark network for RQ2 and RQ3. This choice follows prior studies using Chinese listed-firm top-customer and top-supplier disclosures~\cite{Gao2023SCDisclosureInvestmentEfficiency,Huan2017TopFiveCustomers,Cheng2022SupplierConcentrationInvestors,Xu2023CommonOwnershipSC}. Although other Chinese financial databases support firm-level corporate research, supplier--customer data in this literature generally reflects the same annual-report disclosure boundary. We therefore use CSMAR as a representative disclosure-based benchmark, not as a complete ground truth.

Orbis contains company records for the target firms used in the large-scale experiment, including Shanghai/Shenzhen-listed firms and large unlisted firms as of 2024, and provides firm-name variants for target-firm identification and entity matching~\cite{MoodysOrbis}. Wikidata complements this alias table with multilingual labels and aliases, helping consolidate Chinese legal names, English names, abbreviations, former names, and alternative spellings~\cite{Vrandecic2014Wikidata}.
 
\subsection{Search Configuration}
 
We use Google Search results obtained through the Serper API, which provides programmatic access to search-result fields such as titles, URLs, and snippets~\cite{SerperOfficial}. We issue Chinese-language queries using five templates whose English meanings are \texttt{\{company\_name\}} + ``major suppliers,'' ``key customers,'' ``partners,'' ``buyer supply chain,'' and ``supply-chain partners.'' These templates are designed to capture both direct transaction terms and broader partnership expressions that often appear in public corporate disclosures.

\subsection{Extraction Model}

\begin{figure}[tbp]
  \centering
  \includegraphics[width=\columnwidth]{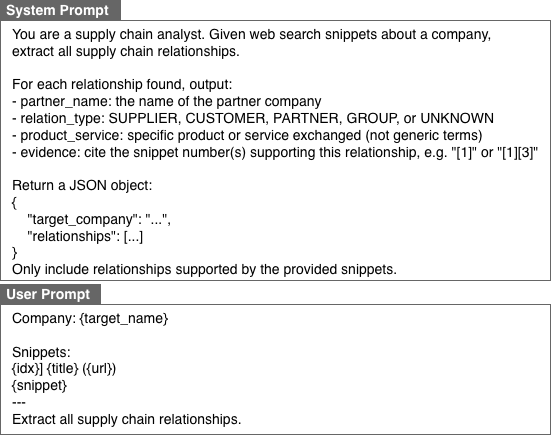}
  \caption{Prompt schema for snippet-based supply-chain relationship extraction. The model receives a target company and numbered search snippets, and returns JSON-formatted relationships with partner names, relation types, exchanged products or services, and supporting snippet identifiers.}
  \label{fig:prompt}
\end{figure}

We use Qwen3-Next-80B-A3B-Instruct as the extraction LLM~\cite{Qwen3NextOfficial}. It is a Qwen3-Next instruction model suitable for long-context structured extraction tasks.
The system prompt instructs the model to act as a supply chain analyst and extract structured relationships from snippets, outputting JSON with fields for partner name, relation type, exchanged product or service, and evidence references. Figure~\ref{fig:prompt} summarizes the prompt schema.

\subsection{Experimental Conditions}

\textbf{Small-scale sampling (RQ1):} 100 Chinese listed companies randomly sampled from the target firm list. We retrieve 30 snippets per query and additionally fetch the full-text content from each snippet's source URL, recording access failures for comparison. For full-text extraction, each successfully fetched source document is split into text chunks before LLM processing because documents may exceed the model input length or contain multiple unrelated sections. Each chunk is processed independently using the same extraction schema, and extracted relationships are deduplicated after entity normalization.

\textbf{Large-scale sampling (RQ2, RQ3):} The large-scale target universe is designed to cover Chinese firms listed on the Shanghai/Shenzhen markets and large unlisted Chinese firms as of 2024. This universe is represented by 130,685 Orbis company records: 4,509 listed firms and 126,176 unlisted firms with operating revenue exceeding USD~130 million. We use these records to identify target firms and obtain firm-name variants, but not as sources of inter-firm edges. The listed subset supports comparison with CSMAR, while the unlisted subset tests expansion beyond listed-firm disclosure. We retrieve 10 snippets per query, reflecting the finding from RQ1 that extraction yield saturates within the top-10 results.

\section{Results}

\subsection{RQ1: Snippet vs.\ Full-Text}

Table~\ref{tab:rq1} compares snippet-based extraction with exhaustive full-text extraction in the 100-firm sample. In the snippet condition, a processed evidence item denotes one retrieved search-result snippet. In the full-text condition, it denotes one successfully fetched and processable source page, whose text is then split into LLM input chunks. Table~\ref{tab:rq1} therefore reports source-level evidence items, while input tokens capture the downstream LLM processing cost after chunking. Rather than serving as a substitute for full-document extraction, snippets provide a lightweight screening layer for broad discovery, while full-text processing can be reserved for selected firms or relationships.

Full-text chunking substantially increases coverage. It yields 41.7$\times$ more raw relationships and 19.8$\times$ more unique relationships than snippets, raising the average number of relationships per target firm from 35.2 to 697.2. This confirms that snippets are query-dependent fragments rather than exhaustive summaries of the underlying documents. However, this coverage gain is costly: full text requires 251.2$\times$ more input tokens, increases total HTTP/API requests from 1,362 to 14,066, and yields much higher input tokens per unique relationship than snippets (7,079 vs.\ 558). The request-level difference arises because both settings require 1,362 Serper API requests for snippet collection, while the full-text setting additionally attempts 12,704 source-page fetches, of which 8,419 yield processable documents.

Full text also produces substantially more duplicate extractions. Its raw-to-unique reduction is 239,181 to 69,024 relationships, corresponding to a duplicate rate of 71.1\%, compared with 39.2\% for snippets. At the same time, relationships per unique partner remain similar across the two settings (1.2 vs.\ 1.1), indicating that the main difference lies in per-firm coverage and repeated extraction volume rather than in the number of distinct relationship records attached to each discovered partner.

\begin{table}[t]
  \centering
  \caption{Comparison of snippet-based and chunked full-text-based SC relationship extraction in the 100-firm sample.}
  \label{tab:rq1}
  \scriptsize
  \begin{tabular}{lrrr}
    \toprule
    \textbf{Metric} & \textbf{Snippet} & \textbf{Full-Text} & \textbf{Full/Snippet} \\
    \midrule
    Total HTTP/API requests            & 1,362        & 14,066        & 10.33$\times$ \\
    Processed evidence items           & 12,704       & 8,419         & 0.66$\times$ \\
    Raw relations                      & 5,741        & 239,181       & 41.66$\times$ \\
    Unique relations                   & 3,488        & 69,024        & 19.79$\times$ \\
    Duplicate rate                     & 39.2\%       & 71.1\%        & +31.9\,pt \\
    Relations per target firm          & 35.2         & 697.2         & 19.81$\times$ \\
    Unique partners                    & 3,181        & 59,111        & 18.58$\times$ \\
    Relations per unique partner       & 1.1          & 1.2           & 1.09$\times$ \\
    Input tokens                       & 1,945,550    & 488,646,427   & 251.16$\times$ \\
    Input tokens per unique relation   & 558          & 7,079         & 12.69$\times$ \\
    \bottomrule
  \end{tabular}
\end{table}

\begin{figure}[tbp]
  \centering
  \includegraphics[width=\columnwidth]{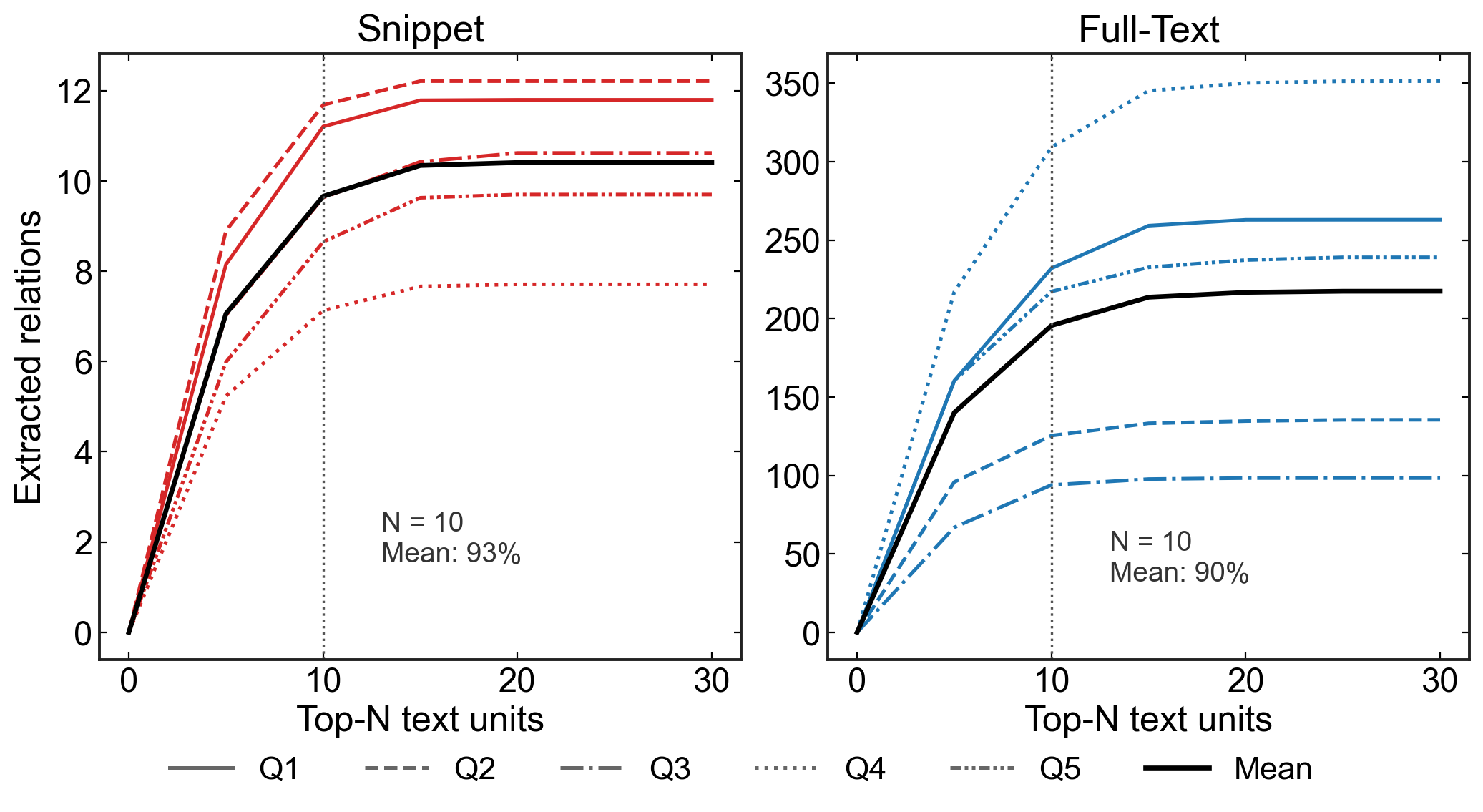}
  \caption{Saturation of extracted supply chain relations as a function of the number of top-ranked search-result items processed per target company and query template. In the snippet condition, an item is processed as a snippet; in the full-text condition, the corresponding source page is fetched and chunked when accessible. Line styles distinguish individual query templates (Q1--Q5); the solid line shows the cross-query mean and shaded bands indicate the min--max range. At $N{=}10$, snippets reach 93\% and full text reaches 90\% of their maximum yields in the 100-firm sample.}
  \label{fig:saturation}
\end{figure}

Figure~\ref{fig:saturation} further shows that extraction yield saturates quickly with respect to the number of top-ranked search-result items processed. Around $N{=}10$, snippets already reach 93\% of their maximum yield, while full text reaches 90\%. This supports the large-scale design used in RQ2: process a shallow but broad set of snippet evidence across many firms, rather than exhaustively processing full text for a much smaller set of firms under the same budget.

Overall, snippets are best interpreted as a broad, low-cost discovery front end, while full text is a high-recall but high-cost option for targeted deep extraction.
 
\subsection{RQ2: Supply Chain Coverage}

The reconstructed domestic inter-firm transaction network provides a large-scale view of China's SC structure. Because CSMAR covers listed-firm disclosures, we use the listed-firm subset of the proposed SCKG for the like-for-like benchmark comparison. This subset captures 7.2$\times$ more unique firms and 9.3$\times$ more transaction relationships than the CSMAR disclosure-based benchmark. The full SCKG covering both listed and unlisted firms further expands coverage to 74.2$\times$ more firms and 110.5$\times$ more relationships, but this reflects a broader target universe rather than a like-for-like comparison.
 
\begin{figure}[tbp]
  \centering
  \includegraphics[width=\columnwidth]{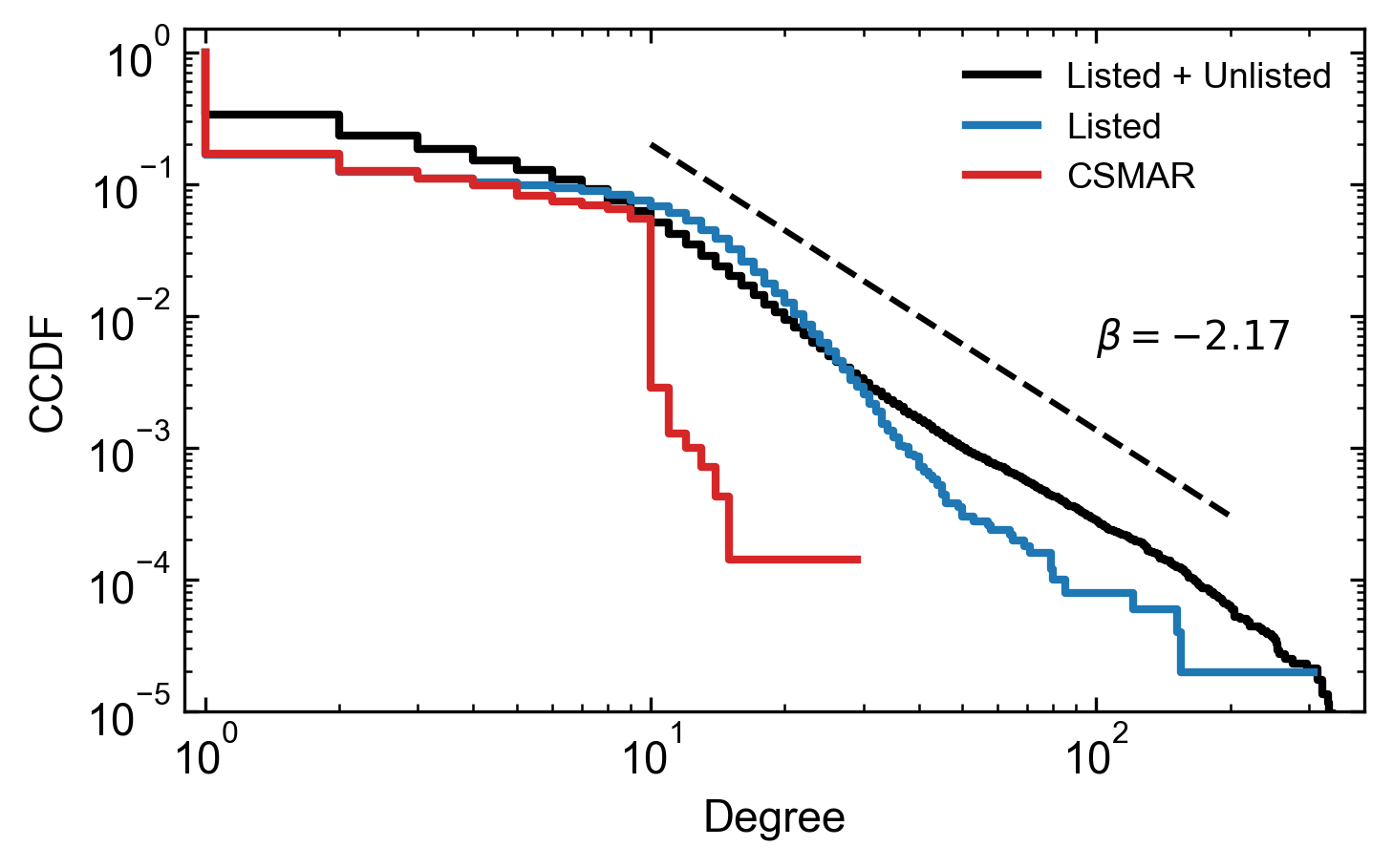}
  \caption{Complementary cumulative distribution function (CCDF) of node degree. The proposed networks exhibit heavy-tailed CCDF patterns. The dashed line shows a fitted slope of $\beta \approx -2.17$ for the listed + unlisted network over $10 \le k \le 200$, where $P(>k) \propto k^{\beta}$. CSMAR shows artificial truncation beyond degree~10 due to disclosure limits.}
  \label{fig:degree_dist}
\end{figure}

Figure~\ref{fig:degree_dist} plots the complementary cumulative distribution function, $P(\mathrm{degree} \geq k)$, on log--log axes. A slower decay indicates the presence of high-degree hub firms. The CSMAR network drops sharply beyond degree~10 because the disclosure regime observes at most five suppliers and five customers per listed firm, fragmenting the network and undermining structural inference. 
In contrast, the snippet-derived networks---both the listed-only subset and the full network covering listed and unlisted firms---retain high-degree hubs and exhibit heavy-tailed CCDF patterns. For the full network, the fitted CCDF slope over $10 \le k \le 200$ is $\beta \approx -2.17$, where $P(>k) \propto k^{\beta}$. These patterns are consistent with empirical regularities observed across firm-level inter-firm transaction networks~\cite{mizuno2014structure,bacilieri2026firm}. This indicates that the proposed method reveals heavy-tailed connectivity patterns in web-visible evidence that are suppressed in disclosure-constrained datasets by construction.

Table~\ref{tab:rq2_top10} compares the top-degree firms in CSMAR, the full proposed SCKG, and the Tier~1-only proposed SCKG. This comparison is not used as the like-for-like coverage benchmark or as an accuracy ranking; rather, it illustrates how different observation windows change the observed hub structure. CSMAR's highest-degree node, State Grid, has only 29 connections, whereas Huawei reaches 1,610 in the full proposed network. This contrast reflects the institutional boundary of listed-firm major-partner disclosures. Because CSMAR is compiled from annual-report disclosures, its network tends to emphasize counterparties that listed firms are more likely to disclose as major customers or suppliers. In the Chinese disclosure setting, such counterparties often include state-owned enterprises, listed firms, infrastructure operators, energy firms, pharmaceutical distribution companies, and resource-related firms. This tendency is visible in the CSMAR degree ranking, where state-owned and disclosure-visible organizations occupy many of the highest-degree positions.

By contrast, several firms that are widely recognized as major industrial actors in China, such as Huawei and BYD, do not appear among the top-ranked CSMAR hubs. This absence does not imply that these firms are unimportant; rather, it reflects the limits of a disclosure-based observation window. The full proposed SCKG instead aggregates web-visible transaction and partnership evidence beyond listed-firm annual reports. As a result, it surfaces industrial hubs across ICT and electronics, EVs and batteries, internet platforms, real estate, and energy. Huawei is particularly informative: although it is not a listed firm in the CSMAR benchmark universe, it appears as the highest-degree node in the full proposed network. This suggests that public web evidence can surface highly visible industrial counterparties that are weakly represented in disclosure-based benchmarks. Other high-degree nodes, including BYD, CATL, China Mobile, Samsung, Tencent, JD.com, and Xiaomi, further indicate that the proposed SCKG captures broader industrial hubs that are expected to be highly visible in public web evidence. The Tier~1-only column provides a stricter credibility-filtered view of the same proposed SCKG, which we discuss further in RQ3.

\begin{table}[t]
  \centering
  \caption{Top-degree firms in CSMAR and the proposed SCKG, with a Tier~1-only view for credibility filtering. Generic or non-firm entities are excluded from the Tier~1-only column.}
  \label{tab:rq2_top10}
  \scriptsize
  \setlength{\tabcolsep}{3pt}
  \begin{tabular}{c l r l r l r}
    \toprule
    Rank &
    \multicolumn{2}{c}{CSMAR} &
    \multicolumn{2}{c}{Proposed All} &
    \multicolumn{2}{c}{Tier~1 only} \\
    \cmidrule(lr){2-3}
    \cmidrule(lr){4-5}
    \cmidrule(lr){6-7}
    & Company & Deg. & Company & Deg. & Company & Deg. \\
    \midrule
    1  & State Grid          & 29 & Huawei       & 1610 & Huawei        & 260 \\
    2  & Sinopharm           & 15 & BYD          & 558  & CNPC          & 135 \\
    3  & Panzhihua Bingyang  & 15 & China Vanke  & 458  & China Railway & 110 \\
    4  & Qilu Huaxin         & 14 & China Mobile & 369  & China Vanke   & 57  \\
    5  & Liandi Info         & 14 & Samsung      & 341  & ICBC          & 53  \\
    6  & Kailuan Energy      & 13 & Tencent      & 333  & PetroChina    & 51  \\
    7  & Shanghai Gold Exch. & 13 & JD.com       & 331  & China Mobile  & 50  \\
    8  & China Energy Inv.   & 12 & Sinopec      & 321  & Sinopec       & 49  \\
    9  & CSG                 & 12 & CATL         & 321  & ZTE           & 48  \\
    10 & Xinhua Pharma       & 11 & Xiaomi       & 313  & BASF          & 47  \\
    \bottomrule
  \end{tabular}
\end{table}

\subsection{RQ3: Credibility Assessment}

To assess whether extracted relationships can be audited after extraction, we apply progressively stricter credibility filters to the snippet-derived network. Table~\ref{tab:rq3_tier} reports network statistics after retaining only relationships whose source domains meet each tier threshold.

\begin{table}[t]
  \centering
  \caption{Network statistics under credibility-tier filtering. Tier~1--$k$ denotes the subgraph obtained by retaining only relationships supported by source domains with credibility Tier~$k$ or better.}
  \label{tab:rq3_tier}
  \begin{tabular}{lrrrr}
    \toprule
    \textbf{Filter} & \textbf{Nodes} & \textbf{Edges} & \textbf{Mean deg.} & \textbf{Max deg.} \\
    \midrule
    \multicolumn{5}{l}{\textit{Listed}} \\
    Tier~1 only       &     476 &     327 & 1.37 &    31 \\
    Tier~1--2         &  11,414 &   9,481 & 1.66 &    64 \\
    Tier~1--3         &  27,670 &  27,790 & 2.01 &   146 \\
    Tier~1--4         &  28,108 &  28,347 & 2.02 &   146 \\
    Tier~1--5 (All)   &  50,600 &  61,086 & 2.41 &   308 \\
    \midrule
    \multicolumn{5}{l}{\textit{Listed + Unlisted}} \\
    Tier~1 only       &   6,015 &   4,957 & 1.65 &   260 \\
    Tier~1--2         & 201,409 & 185,355 & 1.84 &   367 \\
    Tier~1--3         & 260,429 & 273,332 & 2.10 &   456 \\
    Tier~1--4         & 267,665 & 283,528 & 2.12 &   506 \\
    Tier~1--5 (All)   & 522,920 & 722,115 & 2.76 & 1,610 \\
    \midrule
    \multicolumn{5}{l}{\textit{CSMAR}} \\
    All               &   7,051 &   6,534 & 1.85 &    29 \\
    \bottomrule
  \end{tabular}
\end{table}

The results reveal a coverage--credibility trade-off. In the listed-firm subset, filtering from all Tier~1--Tier~5 sources to higher-credibility Tier~1--Tier~3 sources removes 45\% of nodes and 55\% of edges, but the remaining network still contains 27,670 firms and 27,790 edges---3.9$\times$ more firms and 4.3$\times$ more edges than CSMAR. The full network shows the same pattern at a larger scale, retaining 260,429 firms and 273,332 edges.

The Tier~1-only column in Table~\ref{tab:rq2_top10} shows how source filtering changes the observed hubs. Huawei remains the highest-degree firm even under the strictest source filter, suggesting that its centrality is not merely an artifact of low-credibility web sources. Other Tier~1 hubs, such as CNPC, China Railway, ICBC, China Mobile, and BASF, indicate that high-credibility sources retain a recognizable industrial core.

The contrast with the full proposed SCKG reveals a visibility bias in broader web evidence. Consumer-facing or announcement-rich firms such as BYD, Samsung, Tencent, JD.com, CATL, and Xiaomi appear prominently in the full graph, likely because blogs, trade media, corporate news, product announcements, and customer cases make such firms more web-visible. By contrast, the Tier~1-only view shifts toward firms with more official or institutional traces, such as CNPC, China Railway, ICBC, PetroChina, China Mobile, and Sinopec. Credibility tiers should therefore be interpreted not as precision estimates, but as an audit mechanism that exposes the coverage--credibility trade-off and helps prioritize relationships for downstream validation.

\section{Discussion}
\subsection{Key Findings}
 
The findings support interpreting search snippets as a scalable screening layer for web-visible supply-chain evidence. Rather than serving as a substitute for full-document extraction, snippets provide broad first-pass coverage and help identify candidate relationships for subsequent validation. The proposed approach therefore complements disclosure-based financial databases by broadening the observable evidence base beyond formal disclosures.

The comparison with exhaustive full-text extraction clarifies this screening role. Full-text chunking improves per-firm recall, but requires 251.2$\times$ more input tokens and 10.3$\times$ more HTTP/API requests. Snippets therefore sacrifice recall per firm, but make it feasible to expand the target universe from hundreds to over 100,000 firms and to triage where full-text extraction should be applied. Beyond cost efficiency, snippets may also have a salience-filtering effect. Qualitative inspection suggests that query-biased summaries tend to surface publicly prominent or repeatedly mentioned supply-chain relationships, although we do not yet quantify relationship importance or transaction materiality.

When aggregated across many firms, snippet evidence reveals network structures that are consistent with known empirical regularities of inter-firm transaction networks. In the listed-firm comparison, the proposed SCKG contains 7.2$\times$ more firms and 9.3$\times$ more relationships than CSMAR. The full network also exhibits a power-law-like degree distribution consistent with empirical inter-firm trade networks~\cite{mizuno2014structure,bacilieri2026firm}. These results suggest that web-based extraction can surface long-tail relationships and high-degree hubs missed by top-partner disclosure. This does not imply that CSMAR is poorly curated; rather, it reflects the structural boundary of disclosure-based databases, which observe only a subset of economically relevant firm relationships.

These findings also connect the proposed SCKG to the supply-chain opacity problem introduced in the Introduction. In a disclosure-constrained network such as CSMAR, industrially central firms such as Huawei or BYD may appear weakly connected if they are not well captured by listed-firm major-partner disclosures. As a result, analyses based only on such data may understate candidate propagation paths from these firms. By expanding observable counterparties and pathways around such firms, the proposed SCKG provides a broader evidence base for analyzing supply-chain opacity.

The resulting graph remains auditable despite broader extraction. Source-level provenance and credibility tiers make relationships traceable and help prioritize review when exhaustive ground truth is unavailable. Tier-filtered hub rankings show that stricter filters retain recognizable industrial hubs, while broader filters reveal web-visible hubs and sectoral visibility biases. Thus, credibility tiers mainly support review prioritization and sensitivity analysis across source types.

\subsection{Limitations and Future Work}
Several limitations and directions for future work remain. The SCKG should be interpreted as a web-visible evidence graph rather than a complete transaction ledger. Public web evidence is selective and asymmetric: firms may disclose relationships for credibility, marketing, or compliance reasons, whereas confidential, commoditized, or strategically sensitive relationships may remain unobserved. This asymmetry is useful for discovery, but it also means that absence from the graph should not be interpreted as absence of a real relationship. Relationships highlighted by search snippets should also be interpreted as web-visible and potentially salient, rather than necessarily economically material. Assessing transaction size, exposure intensity, or actual shock propagation requires additional validation.

Extraction and normalization remain important areas for improvement. The current pipeline relies on string-based entity matching augmented with alias tables, while product or service fields are extracted as free text. This entity-matching step is a bottleneck: legal names, abbreviations, translations, subsidiaries, and product-specific mentions can split the same firm into multiple nodes or create duplicate relationships. Future work should develop low-cost semantic entity resolution using the snippet context, with vector-based candidate retrieval followed by a compact language model. Product information should also be standardized using controlled taxonomies.

Beyond the Chinese setting, the framework must be adapted to different linguistic and institutional environments. Applying it to other countries or cross-border networks would require revising query templates, credibility-tier rules, and extraction prompts. Increasing query diversity may improve coverage but also raises API and LLM costs. Programmatic optimization frameworks such as DSPy~\cite{Khattab2023DSPy} may help tune these components across domains.

\subsection{Ethical Considerations}
The present study constructs a broad but shallow network from publicly available web information. Should this technology be extended to build deeper corporate intelligence, the potential for weaponization---including competitive intelligence abuse, economic coercion, or targeted sanctions circumvention---must be carefully considered. Responsible disclosure practices and access controls would be essential safeguards.

\section{Conclusion}

We proposed a snippet-driven method for constructing an auditable supply chain knowledge graph (SCKG) from public web search results. The SCKG represents firms as nodes and supply-chain relationships as edges, while retaining provenance metadata for downstream validation. 
The central benefit is scalable first-pass discovery: snippets do not match exhaustive full-text chunking in relationship recall, but they avoid making full-text fetching and document-scale LLM processing prerequisites for every firm. In experiments on Chinese supply chains, full-text processing extracted 19.8$\times$ more unique relationships than snippets, but required 251.2$\times$ more input tokens and 10.3$\times$ more HTTP/API requests. At scale, the snippet-driven pipeline was applied to 130,685 target firms. In the listed-firm subset, it produced an SCKG covering 7.2$\times$ more firms and 9.3$\times$ more relationships than the China Stock Market \& Accounting Research Database (CSMAR) disclosure-based benchmark while preserving provenance for downstream review.

These findings position snippet-driven extraction as a complement to disclosure-based databases, not a replacement. CSMAR remains valuable as structured, research-ready information on listed Chinese firms, but its coverage is constrained by the underlying disclosure regime: firms disclose only major partners, and some records may be anonymized or insufficiently identifiable. 
By expanding beyond this boundary, the proposed SCKG makes industrial hubs and candidate propagation pathways more visible than in disclosure-based networks alone. It should therefore be interpreted as a web-visible evidence layer that broadens the observable basis for supply-chain risk analysis while preserving provenance for subsequent validation.

\section{Acknowledgement}
This work was supported by JSPS KAKENHI Grant Numbers JP25K01458 and JP23H00042, and JST A-STEP Grant Number JPMJTR25RL.

\bibliographystyle{IEEEtran}
\bibliography{library}

\end{document}